\documentclass{PoS}

\title{Exclusive diffractive photon bremsstrahlung at high energies }

\ShortTitle{Exclusive diffractive photon bremsstrahlung at high energies }

\author{Piotr LEBIEDOWICZ%
\thanks{This work was supported by the Polish National Science Centre 
(on the basis of decision No.~DEC-2011/01/N/ST2/04116
and DEC-2011/01/B/ST2/04535).}\\
Institute of Nuclear Physics PAN, PL-31-342 Cracow, Poland\\
E-mail: \email{Piotr.Lebiedowicz@ifj.edu.pl}} 
\author{\speaker{Antoni SZCZUREK}\\
University of Rzesz\'ow, PL-35-959 Rzesz\'ow, Poland, and\\
Institute of Nuclear Physics PAN, PL-31-342 Cracow, Poland\\
E-mail: \email{Antoni.Szczurek@ifj.edu.pl}}

\abstract{
We discuss the $p p \to p p \gamma$ reaction at the LHC.
We consider diffractive classical bremsstrahlung mechanisms including effects
of non point-like nature of protons.
In addition, we take into account (vector meson)-pomeron, photon-pion
as well as photon-pomeron exchange processes for the first time in the literature.
The integrated diffractive bremsstrahlung cross section
($E_{\gamma}>100$~GeV) is only of the order of $\mu$b.
We try to identify regions of the phase space where one of the mechanisms dominates.
The classical bremsstrahlung dominates 
at large forward/backward photon pseudorapidities, 
close to the pseudorapidities of scattered protons.
In contrast, the photon-pomeron (pomeron-photon) mechanism
dominates at midrapidities but the related cross section is rather small. 
In comparison the virtual-omega--rescattering mechanism contributes at smaller angles of photons 
(larger photon rapidities).
}

\FullConference{XXI International Workshop on Deep-Inelastic Scattering and Related Subject -DIS2013,\\
		22-26 April 2013\\
		Marseilles,France}

\begin{document}

%---------------------------
\section{Introduction}
%---------------------------
Exclusive diffractive photon bremsstrahlung mechanism have been studied
recently \cite{LS13_gamma} in the $p p \to p p \gamma$ reaction at the LHC.
%has been studied the exclusive $p p \to p p \gamma$ reaction at the LHC.
\footnote{
The $p p \to p p \gamma$ process at high energies was discussed before in
\cite{Khoze} and it was proposed to use the exclusive photon bremsstrahlung 
to measure or estimate elastic proton-proton cross section at the LHC.
But only the approximate formulas for the classical bremsstrahlung were given there.
The participating particles were treated there as point-like particles.
No differential distributions for the exclusive bremsstrahlung have been discussed.}
Because at high energy the pomeron exchange is the driving mechanism of bremsstrahlung 
it is logical to call the mechanisms 
described by the diagrams from (a) to (d) in Fig.\ref{fig:diagrams}
the diffractive bremsstrahlung to distinguish from the low-energy
bremsstrahlung driven by meson exchanges.

We shall include bremsstrahlung diagrams 
as well as some new diagrams characteristic exclusively for
proton-proton scattering, not present e.g. in $e^+ e^-$ scattering.
We include diagrams which arise in the vector-dominance model as well as
photon-pion (pion-photon) and photon-pomeron (pomeron-photon)
exchange processes not discussed so far in the literature.
We shall try to identify the region of the phase space where one 
can expect a dominance of one of the processes
through detailed studies of several differential distributions.
%\footnote{The photon bremsstrahlung was intensively studied
%in nucleon-nucleon collisions at low energies 
%(see e.g. \cite{Nakayama,Scholten} and references therein).
%a beam energy of 310 MeV .
%There the dominant mechanisms are nucleon current (off-shell nucleon) 
%and/or mesonic current (photon emitted from the middle of exchanged mesons)
%contributions driven by meson exchanges.}.
The exclusive photon production mechanism
is similar to $p p \to p p \omega$ \cite{CLSS} and $p p \to p p \pi^{0}$ \cite{LS13_pi0} processes.
As discussed in the past the dominant hadronic bremsstrahlung-type 
mechanism is the Drell-Hiida-Deck mechanism \cite{Deck}
for diffractive production of $\pi N$ final states
(for a nice review we refer to \cite{reports} and references therein).
The photons radiated off the initial and final state protons 
can be seen by the Zero Degree Calorimeters (ZDCs) \cite{ZDC}
%that are installed at about $140$ meters on each side of the interaction region.
%They will measure very forward neutral particles
in the pseudorapidity region $|\eta|>8.5$ and $8.3$ at the CMS and ATLAS, respectively.
\footnote{The exclusive $pp \to nn \pi^{+}\pi^{+}$ \cite{LS11}
can be also measured with the help of the ZDCs.
Very large cross sections has been found which
is partially due to interference of a few mechanisms.}
The scattered protons can be tagged e.g. by the ALFA detectors at ATLAS (see \cite{SLTCS11})
or TOTEM at CMS.

%----------------------------------------------------
\section{Sketch of formalism}
%----------------------------------------------------
%-----------------------------------------------------------------
\begin{figure}[!ht]
%\centering
(a)\includegraphics[width=3.5cm]{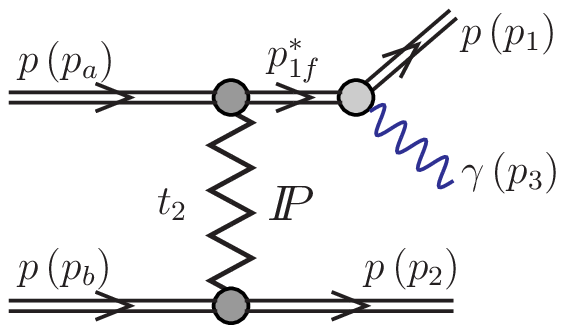}
(b)\includegraphics[width=3.5cm]{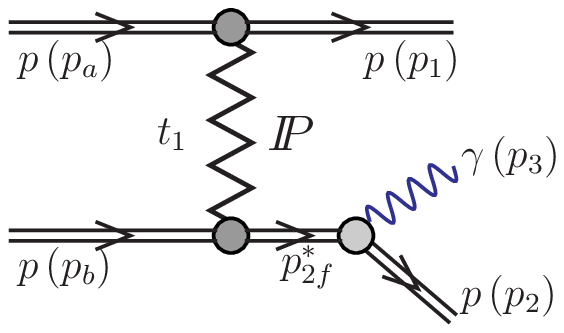}
(c)\includegraphics[width=3.cm]{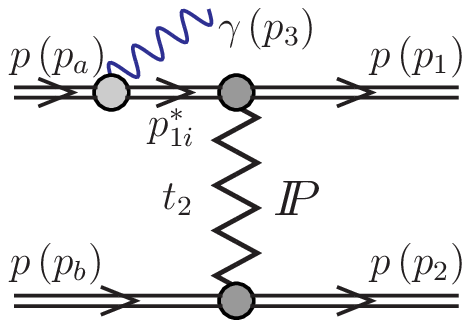}
(d)\includegraphics[width=3.cm]{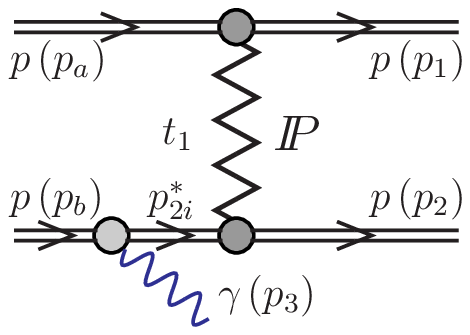}
(e)\includegraphics[width=3.cm]{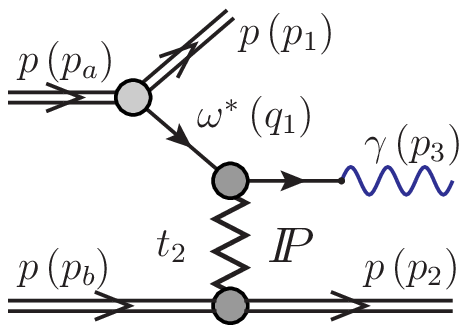}
(f)\includegraphics[width=3.cm]{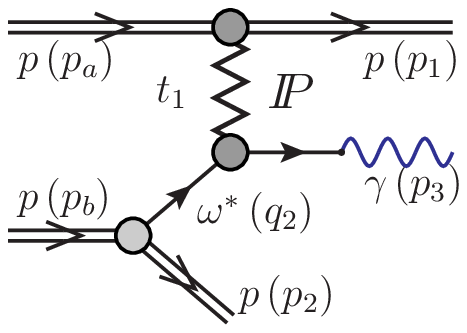}
(g)\includegraphics[width=3.cm]{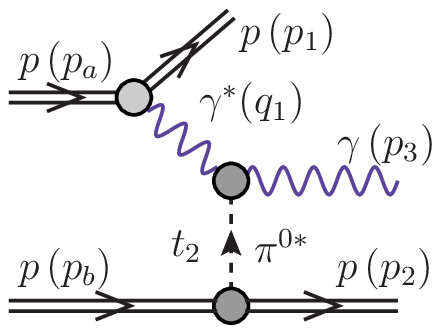}
(h)\includegraphics[width=3.cm]{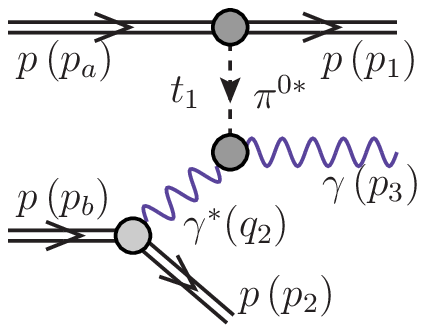}
(i)\includegraphics[width=3.cm]{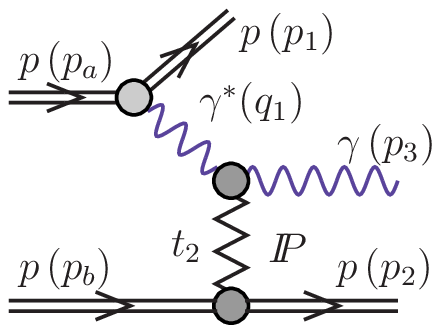}
(j)\includegraphics[width=3.cm]{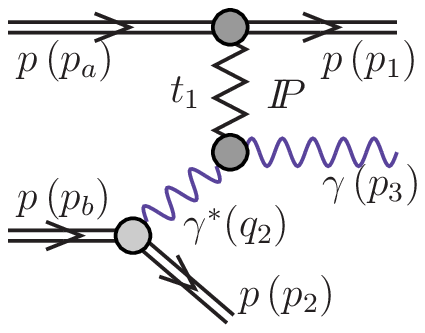}
\caption{\label{fig:diagrams}
\small Diagrams of the dominant diffractive bremsstrahlung (a) - (d)
driven by the pomeron exchange, and from (e) to (j), the virtual $\omega$ - pomeron exchanges
(or $\omega$-rescattering), the $\gamma$ - $\pi^{0}$ exchanges,
and the $\gamma$ - pomeron exchanges (or $\gamma$-rescattering).
}
\end{figure}
%------------------------------------------------------------------
The amplitudes for exclusive production of photons shown schematically 
by diagrams in Fig.\ref{fig:diagrams} are discussed elsewhere \cite{LS13_gamma},
e.g. formula of diagram (a) can be written as
%It is straightforward to evaluate the contribution read:
%
\begin{eqnarray}
{\cal M}^{(a)}_{\lambda_{a}\lambda_{b} \to \lambda_{1}\lambda_{2}\lambda_{3}} &=&
e\;
\bar{u}(p_{1},\lambda_{1}) 
%\Gamma^{\mu}_{\gamma N N^{*}}(p_{3})
\varepsilon^{*}\!\!\!\!\!\!\!\!/ \,(p_{3},\lambda_{3})
S_{N}(p_{1f}^{2})
\gamma^{\mu}
u(p_{a},\lambda_{a})\;
%\varepsilon_{\mu}^{*}(p_{3},\lambda_{3}) \;
F_{\gamma N^{*}N}(p_{1f}^{2}) \;
F_{I\!\!P NN^{*}}(p_{1f}^{2}) \nonumber \\
&&\times 
i s \,C_{I\!\!P}^{NN} \left( \frac{s}{s_{0}}\right)^{\alpha_{I\!\!P}(t_{2})-1} 
\exp\left(\frac{B_{I\!\!P}^{NN} t_{2}}{2}\right)\,\frac{1}{2s}\;
%\delta_{\lambda_{2}\lambda_{b}},
\bar{u}(p_{2},\lambda_{2}) 
\gamma_{\mu} 
u(p_{b},\lambda_{b})\,.
\label{brem_a}
\end{eqnarray}
We use interaction parameters of Donnachie-Landshoff \cite{DL92}
fixed to the total and elastic $\pi N$ and $NN$ scattering, see \cite{LS10}.
In the bremsstrahlung processes discussed here the intermediate nucleons are off-mass shell.
In our approach the off-shell effects related
to the non-point-like protons in the intermediate state are included by 
the following simple extra form factors
\begin{eqnarray}
F(p^{2}) = \frac{\Lambda_{N}^{4}}{(p^{2}-m_{p}^{2})^{2} + \Lambda_{N}^{4}}\,.
\label{extra_ff}
\end{eqnarray}
%This form was used e.g. in Ref.\cite{OTL01} for $\omega$ photoproduction.
%In general, 
The cut-off parameters $\Lambda_{N}$ in the form factors are not known
but could be fitted in the future to the (normalized) experimental data.
%From our general experience in hadronic physics we expect $\Lambda_{N} \sim 1$~GeV.
We shall discuss how the uncertainties of the form
factors influence our final results.

%----------------------------------------------------
\section{Results}
%----------------------------------------------------
In the following section we show results of 
the differential distributions for the all mechanisms of Fig.\ref{fig:diagrams}.
In calculating cross section we perform integrations in
$\xi_1 = \log_{10}(p_{1\perp}/1\,\mathrm{GeV})$ %(for $\gamma \pi$-exchange)
and $\xi_2 = \log_{10}(p_{2\perp}/1\,\mathrm{GeV})$ %(for $\pi \gamma$-exchange)
instead in $p_{1\perp}$ and $p_{2\perp}$, 
in the photon pseudorapidity $\eta_{\gamma}$
and the relative azimuthal angle between the outgoing protons
$\phi_{12} = \phi_{1} - \phi_{2}$.

\begin{figure}[!ht]
\centering
\includegraphics[width = 0.49\textwidth]{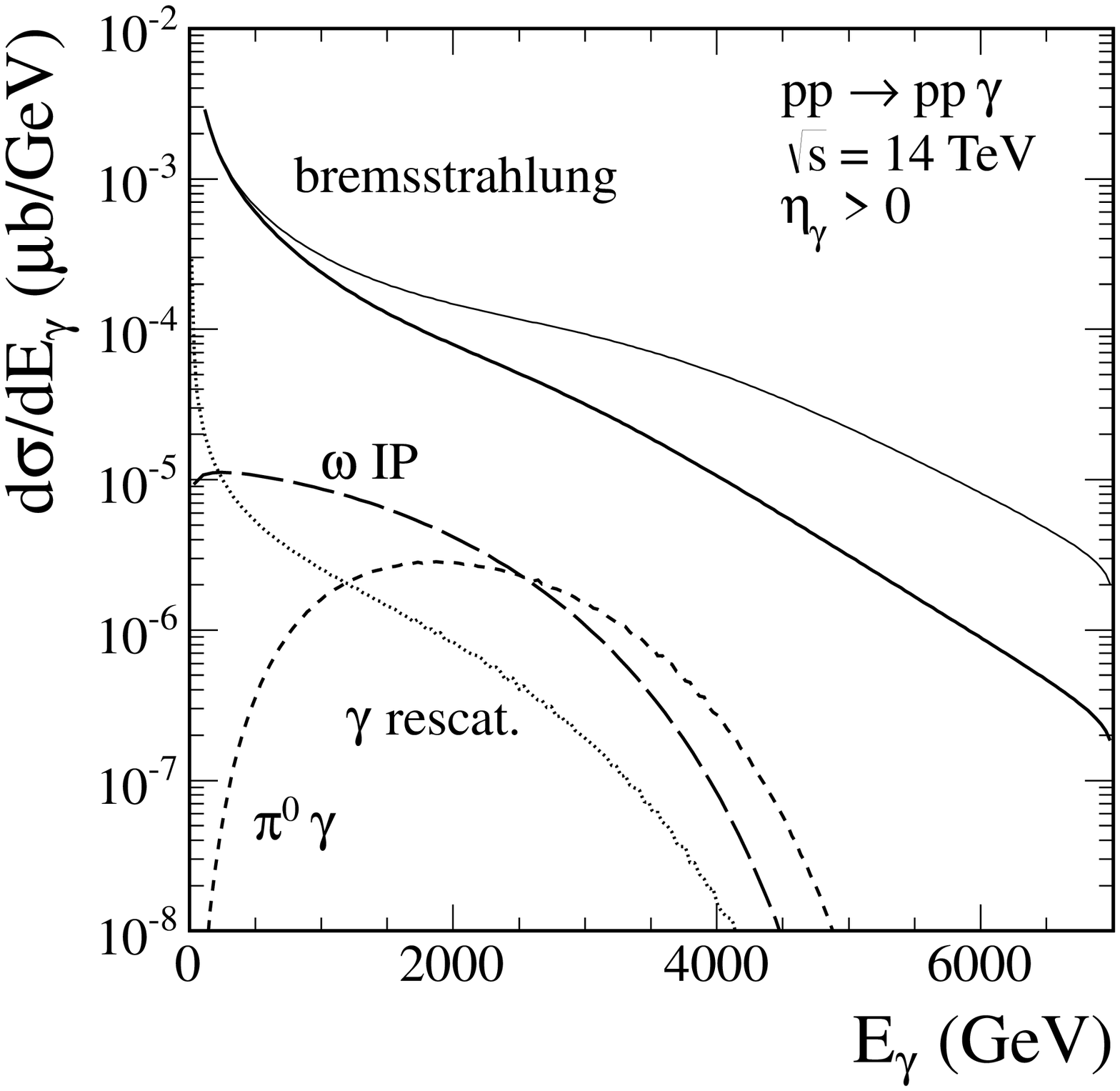}
\includegraphics[width = 0.49\textwidth]{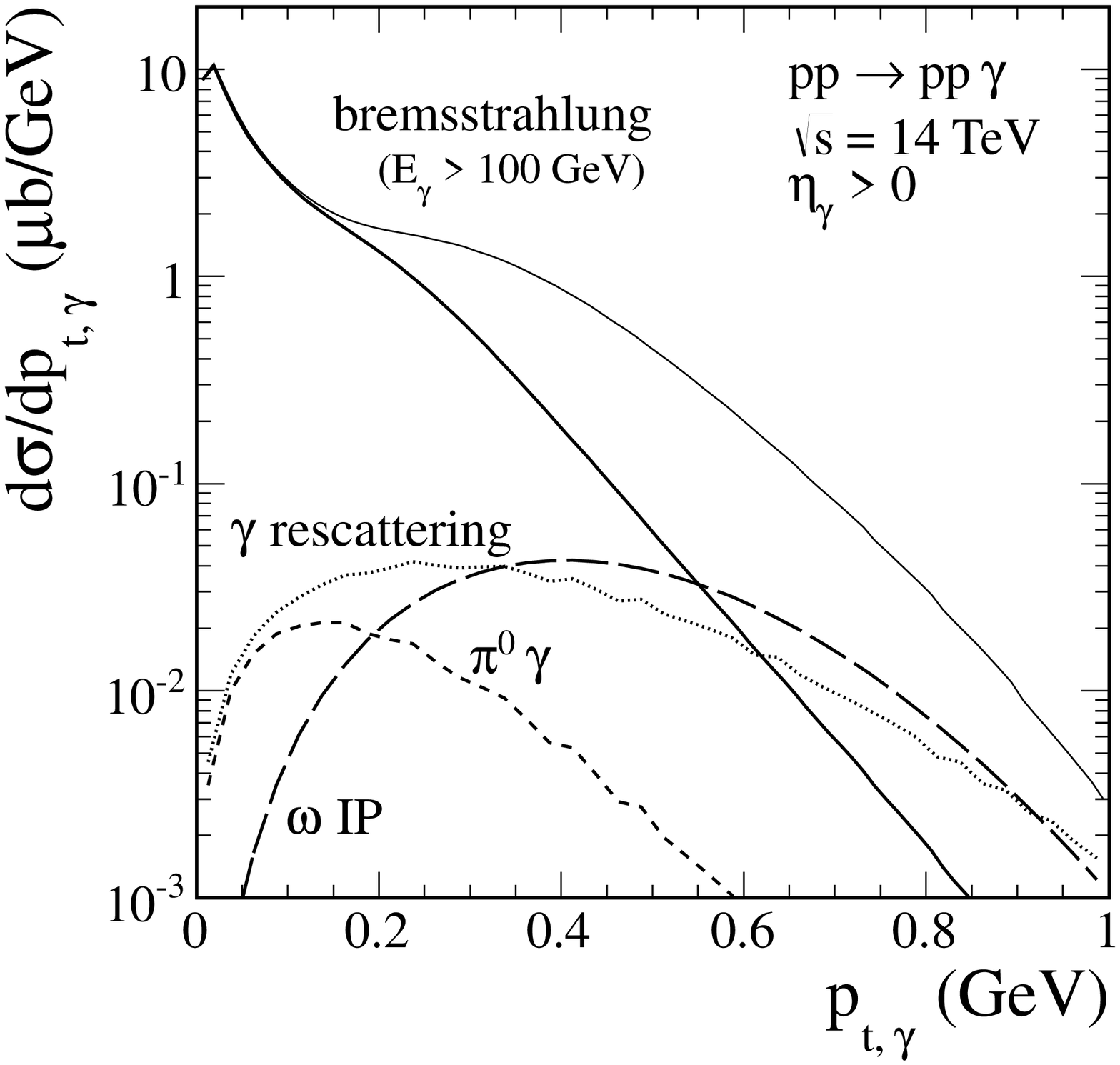}
  \caption{\label{fig:brem_dsig_dE3}
Energy spectrum of photons (left panel)
and distribution in transverse momentum of photons (right panel)
for all processes considered here at $\sqrt{s} = 14$~TeV and for $\eta_{\gamma}>0$.
For classical bremsstrahlung we have imposed $E_{\gamma} > 100$~GeV
and used two values of $\Lambda_{N} = 0.8, 1$~GeV 
in the proton off-shell form factors
% (Eq.(\ref{extra_ff}))
(see the lower and upper solid line, respectively).
}
 \end{figure}

Corresponding distributions in the photon energy and the transverse momentum are 
shown in Fig.\ref{fig:brem_dsig_dE3}. 
The ZDC detectors can measure only photons above some energy threshold.
In the calculation of classical bremsstrahlung
presented here we assume $E_{\gamma} > 100$~GeV as an example.
The contribution of classical bremsstrahlung is concentrated at very small transverse momenta
which is consistent with very small photon emission angle (large
pseudorapidity). The other distributions have rather similar shape
and vanish at $p_{\perp,\gamma} = 0$~GeV.
The exact shape may depend somewhat on the functional form
and values of cut-off parameters of off-shell
form factors taking into account the non-point-like nature of the vertices involved.
Here we have fixed the values of the corresponding form factors at typical hadronic scales.
%--------------------------------------------------------
%--------------------------------------------------------

\begin{figure}[!ht]
\centering
\includegraphics[width = 7.cm]{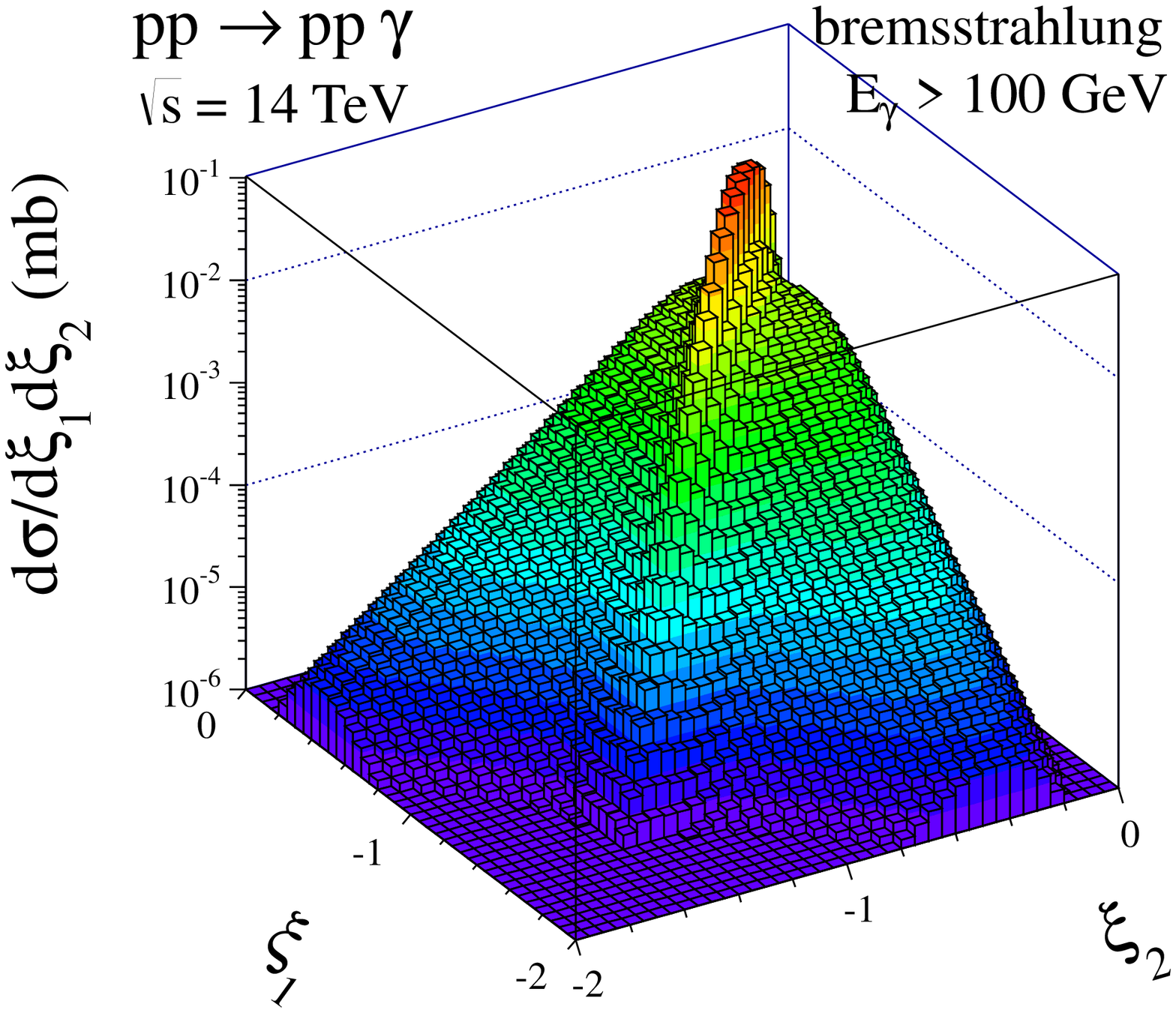}
\includegraphics[width = 7.cm]{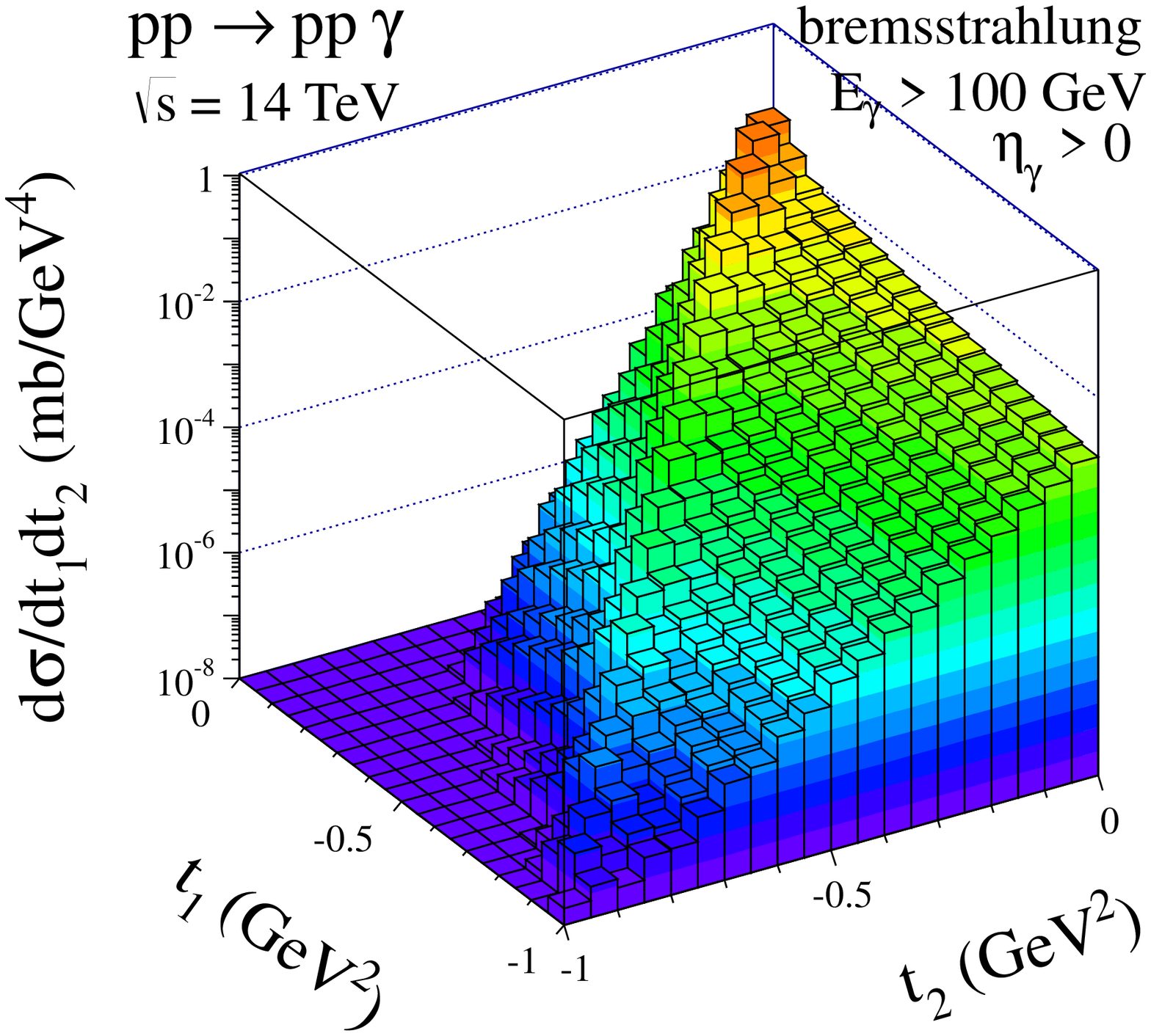}
%(b)\includegraphics[width = 7.cm]{map_xi1xi2_ef.eps}\\
%(c)\includegraphics[width = 7.cm]{map_xi1xi2_gh.eps}
%(d)\includegraphics[width = 7.cm]{map_xi1xi2_ij.eps}
  \caption{\label{fig:maps}
  \small
Distribution in 
$(\xi_1,\xi_2)$ = $(\log_{10}(p_{1\perp}/1\,\mathrm{GeV}), \log_{10}(p_{2\perp}/1\,\mathrm{GeV}))$
(left panel) and in ($t_1, t_2$) (right panel)
for the classical bremsstrahlung mechanisms at $\sqrt{s} = 14$~TeV.
We have imposed in addition $E_{\gamma} > 100$~GeV and used $\Lambda_{N} = 1$~GeV.
}
\end{figure}

In Fig.\ref{fig:maps} we show two-dimensional distributions
for the classical bremsstrahlung
in $(\xi_1,\xi_2)$ in a full range of photon (pseudo)rapidity
and in $(t_1,t_2)$ for $\eta_{\gamma}>0$.
We observe an enhancement along the diagonal. 
This enhancement is a reminiscence of the elastic scattering
for which $\xi_1 = \xi_2$ or $t_{1} = t_{2}$.
The distributions discussed here could in principle be obtained with 
the TOTEM detector at CMS to supplement the ZDC detector for the measurement of photons.
%--------------------------------------------------------
%--------------------------------------------------------

\begin{figure}[!ht]
\centering
\includegraphics[width = 0.49\textwidth]{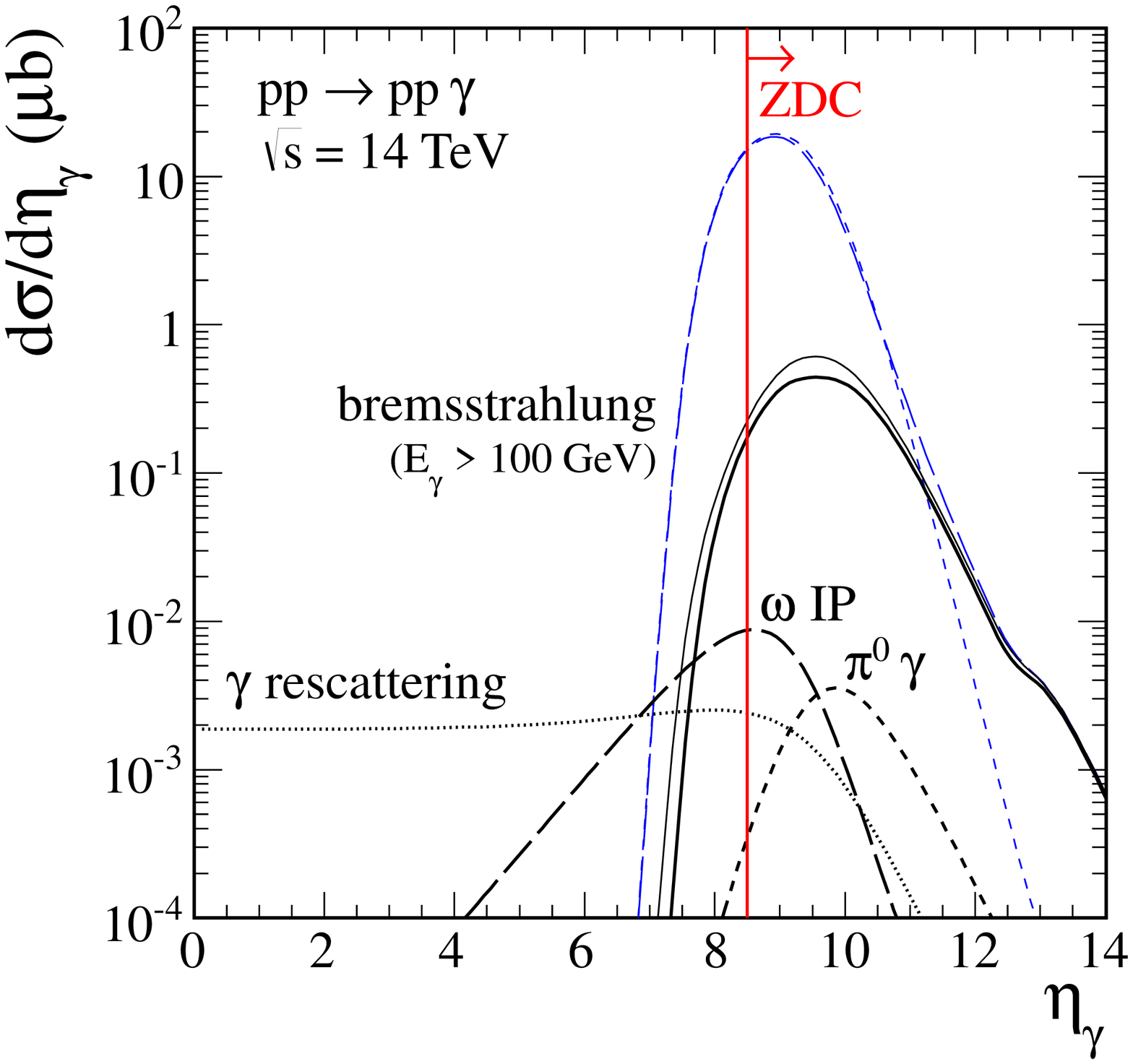}
\includegraphics[width = 0.49\textwidth]{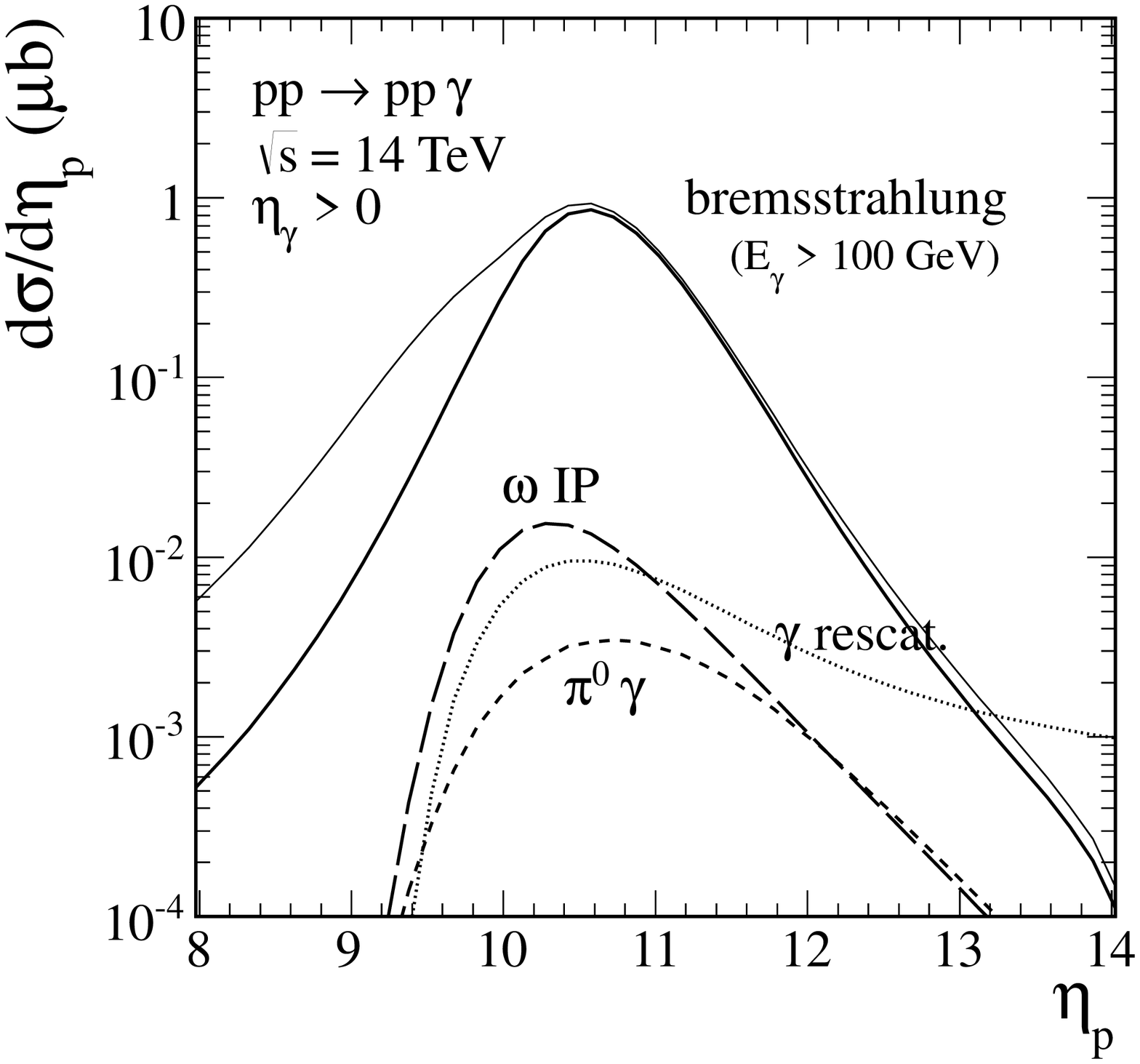}
  \caption{\label{fig:dsig_dy}
  \small
Distribution in (pseudo)rapidity of emitted photons (left panel) and 
in pseudorapidity of outgoing protons (right panel) at $\sqrt{s} = 14$~TeV.
For classical bremsstrahlung we have imposed $E_{\gamma} > 100$~GeV
and used two values of $\Lambda_{N} = 0.8, 1$~GeV
(the lower and upper solid line, respectively).
%A large cancellation between the initial (\ref{brem_c}) and final state radiation (\ref{brem_a})
%is shown (see the blue short-dashed and the blue long-dashed lines, respectively).
The lower pseudorapidity limit for the CMS ZDC detector ($\eta_{\gamma} > 8.5$)
is shown in addition by the vertical line.
%The vertical line indicate the lower pseudorapidity limit for the CMS ZDC detector.
}
\end{figure}

Photon (pseudo)rapidity distribution is particularly interesting.
In Fig.\ref{fig:dsig_dy} we show both distributions for photons $\eta_{\gamma}$ (left panel)
and corresponding distributions for outgoing protons $\eta_{p}$ (right panel).
In this variable both protons and photons 
are localized in a similar region of pseudorapidities (or equivalently polar angles).
%In Fig.\ref{fig:dsig_dy} we compare distribution in photon
%pseudorapidity for all processes considered in the present paper.
The classical bremsstrahlung clearly gives the largest contribution.
It is also concentrated at very large $\eta_{\gamma}$ 
i.e.~in the region where ZDC detectors can be used.
We observe a large cancellation between the corresponding terms in the amplitude
(as shown in the left panel between diagrams (a) and (c)
denoted by the blue long-dashed and the blue short-dashed lines, respectively).
%\textbf{The omega-rescattering and pion cloud mechanisms contribute to the region of somewhat
%smaller pseudorapidities. 
%This is more difficult region as far as experimental measurements are considered.}
The $\gamma$-rescattering process with pomeron exchange 
clearly dominates in the region of $\eta_{\gamma} < 6$.
The cross section for this process is rather small.
Clearly an experimental measurement there would be a challenge.
%-------------------------------------------------------------------------
%-------------------------------------------------------------------------

\begin{figure}[!ht]
\centering
\includegraphics[width = 0.49\textwidth]{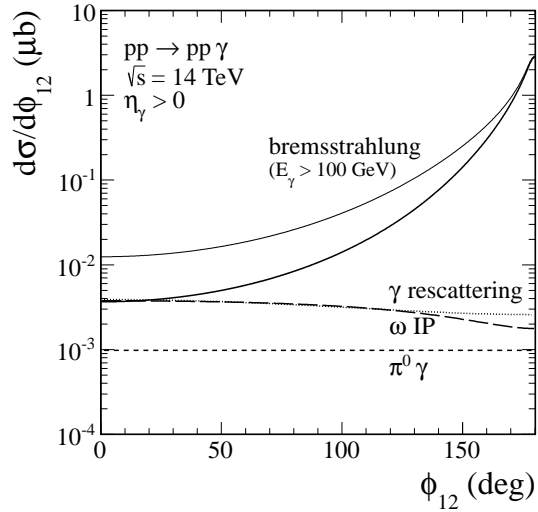}
  \caption{\label{fig:dsig_dphi}
  \small
Distribution in relative azimuthal angle between outgoing protons
for different mechanisms at $\sqrt{s} = 14$~TeV and for $\eta_{\gamma}>0$.
For classical bremsstrahlung we have imposed $E_{\gamma} > 100$~GeV
and used two values of $\Lambda_{N} = 0.8, 1$~GeV 
(the lower and upper solid line, respectively).
}
 \end{figure}

If both protons are measured one could also study correlations
in the relative azimuthal angle between outgoing protons. 
Our model calculations are shown in Fig.\ref{fig:dsig_dphi}. 
One can observe a large enhancement at back-to-back configurations 
for the classical bremsstrahlung which reminds the elastic scattering case 
($\phi_{12} = \pi$).
The contributions of other mechanisms are significantly smaller
and weakly depend on $\phi_{12}$.
%--------------------------------------------------------
%--------------------------------------------------------

%--------------------------
\section{Conclusions}
%--------------------------

We have considered several mechanisms of exclusive single photon production
and calculated differential distributions at the LHC nominal energy.
By imposing several cuts one could select or enhance the contribution of one of the mechanisms. 
The classical bremsstrahlung mechanism turned out to give the biggest
cross section concentrated at large photon (pseudo)rapidities. 
The photons are emitted at only slightly smaller pseudorapidities than the scattered protons.
%The photons are emitted at similar angles as outgoing protons.
We observe a strong cancellation between the initial and final state radiation.
The cross section for the classical bremsstrahlung is peaked at 
back-to-back configurations (similar transverse momenta or polar angles of 
outgoing protons and relative azimuthal angle concentrated close to $\phi_{12} = \pi$).
This is a clear reminiscence of elastic scattering. 
Cut on photon energy ($E_{\gamma} > 100$~GeV) reduces 
the region of $\phi_{12} \cong \pi$ significantly
and the integrated diffractive bremsstrahlung cross section
is only of the order of $\mu$b.
The cross section for pion-photon or photon-pion exchanges is much smaller.
Here both small (photon exchange) and large (pion exchange)
four-momentum transfers squared are possible. 
For this process there is no correlation in azimuthal angle between outgoing protons.
The classical bremsstrahlung mechanisms
could be studied in a close future 
with the help of Zero Degree Calorimeters (photons)
and the ALFA or TOTEM detectors (protons).
%The cross section for pomeron-photon or photon-pomeron exchanges is rather small
%and concentrated at midrapidities. 
%Furthermore, the transverse momenta of outgoing 
%photons are small and cannot be easily measured with central ATLAS or CMS detectors.

Summarizing, even present LHC equipment allows to study
exclusive production of photons. Since this process was never
studied at high energies it is worth to make efforts to obtain
first experimental cross sections. 
Since the cross sections are reasonably large 
one could try to obtain even some differential distributions.
This would allow to test our understanding of the diffractive processes 
and help in pinning down some hadronic and electromagnetic off-shell form factors, 
difficult to test otherwise.

\end{document}